\documentclass[conference]{IEEEtran}
\IEEEoverridecommandlockouts
\usepackage{cite}
\usepackage{amsmath,amssymb,amsfonts}
\usepackage{algorithmic}
\usepackage{graphicx}
\usepackage{textcomp}
\usepackage{xcolor}

\usepackage[hyphens]{url}
\usepackage{subfigure}

\def\BibTeX{{\rm B\kern-.05em{\sc i\kern-.025em b}\kern-.08em
    T\kern-.1667em\lower.7ex\hbox{E}\kern-.125emX}}
\begin{document}

\title{Anableps: Adapting Bitrate for Real-Time Communication Using VBR-encoded Video
\thanks{This work is partially supported by the National Natural Science Foundation of China (62101241), Jiangsu Provincial Double-Innovation Doctor Program (JSSCBS20210001), and ZTE Collaborative Research fund. {\it (Corresponding Authors: Hao Chen.)}}
}

\author{\IEEEauthorblockN{Zicheng Zhang}
\IEEEauthorblockA{\textit{Nanjing University}\\
Nanjing, China \\
zichengzhang@smail.nju.edu.cn}
\and
\IEEEauthorblockN{Hao Chen}
\IEEEauthorblockA{\textit{Nanjing University}\\
Nanjing, China \\
chenhao1210@nju.edu.cn}
\and
\IEEEauthorblockN{Xun Cao}
\IEEEauthorblockA{\textit{Nanjing University}\\
Nanjing, China \\
caoxun@nju.edu.cn}
\and
\IEEEauthorblockN{Zhan Ma}
\IEEEauthorblockA{\textit{Nanjing University}\\
Nanjing, China \\
mazhan@nju.edu.cn}
}

\maketitle
\begin{abstract}
Content providers increasingly replace traditional constant bitrate with variable bitrate (VBR) encoding in real-time video communication systems for better video quality. However, VBR encoding often leads to large and frequent bitrate fluctuation, inevitably deteriorating the efficiency of existing adaptive bitrate (ABR) methods. To tackle it, we propose the \textit{Anableps} to consider the network dynamics and VBR-encoding-induced video bitrate fluctuations jointly for deploying the best ABR policy. With this aim, \textit{Anableps} uses sender-side information from the past to predict the video bitrate range of upcoming frames. Such bitrate range is then combined with the receiver-side observations to set the proper bitrate target for video encoding using a reinforcement-learning-based ABR model. As revealed by extensive experiments on a real-world trace-driven testbed, our \textit{Anableps} outperforms the {\it de facto} GCC  with significant improvement of quality of experience, e.g., 1.88$\times$ video quality, 57\% less bitrate consumption, 85\% less stalling, and 74\% shorter interaction delay.
\end{abstract}


\section{Introduction}
Real-time video communication (RTVC) applications like video conferencing, cloud gaming, metaverse, etc, grow rapidly in the past years~\cite{COVID-19}. Different from conventional Video on Demand (VoD) services, the client in the RTVC system usually cannot maintain a multi-second buffer to amortize the short-term network variations because of the low latency constraint. 
In order to guarantee the quality of experience (QoE) 
under heterogeneous networks (e.g., Cellular, WiFi, Starlink, etc.), adaptive bitrate (ABR) technology is widely used to combat network dynamics.

Nowadays, most content providers prefer the VBR mode for video encoding~\cite{huang2014buffer, reed2016leaky
}, as it can achieve better visual quality at the same bitrate budget when compared with the constant bitrate (CBR) encoding~\cite{lakshman1998vbr}. This is because the VBR encoding can intelligently allocate bits according to the frame characteristics. For example, more bits are allocated for frames with complex intensive scenes and fewer for simple stationary scenes, which leads to the improvement of video quality but unfortunately yields a large bitrate fluctuation from time to time. 

\begin{figure}[t]
\setlength{\abovecaptionskip}{0.cm}
\centering
\includegraphics[width=0.48\textwidth]{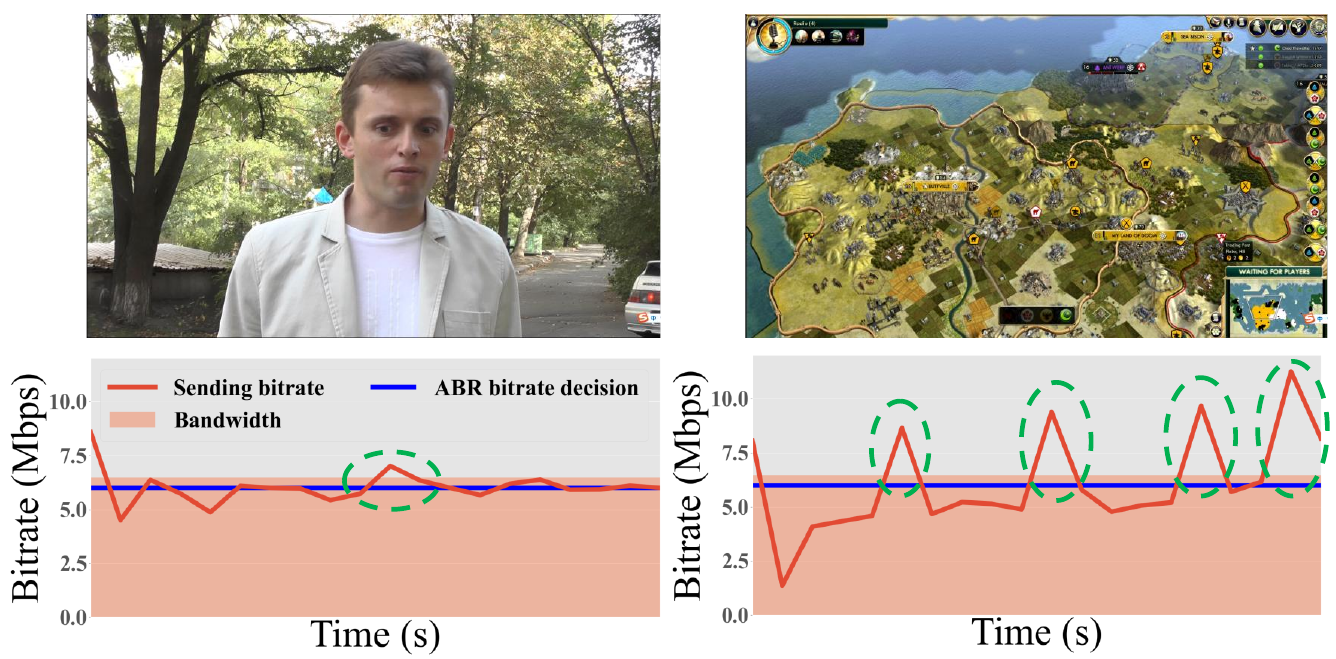}
\caption{{\bf VBR in RTVC}. VBR-encoding-induced video bitrate fluctuates in typical UGC-based live broadcasting scenarios.
The overshooting events in green-dotted eclipses occur when the instant sending bitrate exceeds the available bandwidth.}
\label{fig:all}
\vspace{-0.5cm}
\end{figure}

{\bf Observations In-the-Wild.} As for most existing ABR algorithms used in RTVC systems, they are tuned to maximize bandwidth utilization by increasing the sending bitrate before the overshooting occurs. Apparently, such a VBR-encoding-induced large bitrate fluctuation inevitably raises a challenge that the actual sending bitrate easily exceeds the available bandwidth when attempting to execute the ABR bitrate decision by these ABR algorithms, thus deteriorating the QoE. This phenomenon is more visible in UGC (user-generated content) services like E-commerce or gaming live broadcasting. 

As shown in Fig.~\ref{fig:all}, two UGC videos (one is the live interview with the static scenes, and the other is gaming broadcasting with the dynamic scenes) were encoded at 6Mbps using the VBR mode and then respectively transmitted over the network with a fixed bandwidth of 6.5Mbps. Both sessions encountered overshooting events, leading to video stalling, playback delay, and QoE degradation. The bitrate fluctuation when encoding the dynamic-scene video on the right in VBR mode was more severe, yielding more frequent overshooting.

{\bf Our Approach.} To this end, we propose the \textit{Anableps}\footnote{Anableps is a species of four-eyed fish that can see the objects above and below the water at the same time, which is more or less the same as the proposed approach that learns bitrate decision policies by observing both network dynamics and video content related bitrate fluctuations.}, a learning-based ABR approach used in popular RTVC systems to best optimize the delivery of VBR-encoded videos for better QoE. \textit{Anableps} has two basic functions: the CBPN (Compressed Bitrate Prediction Network) which is used to predict the bitrate range of upcoming video frames according to the sender-side video information (i.e., content complexity and encoding setting) in the past, and the ABRN (Adaptive BitRate Network) which makes bitrate decisions based on the video bitrate range from the CBPN, network conditions (i.e., sending bitrate, receiving bitrate, packet loss rate, round-trip time, and negative acknowledgment count), and receiver-side playback status (i.e., average frame delay and lost frame rate). 

{\bf Contribution.} 
(1) To the best of our knowledge, this work is the first one to consider the impact of video bitrate fluctuations on the ABR decisions in the existing RTVC system, especially for VBR-encoded videos. 
(2) To capture the video bitrate fluctuations, we propose to predict the video bitrate range of upcoming frames in real time using sender-side information like spatiotemporal indices, I frame indicators and bitrate variations. Such a bitrate range is then combined with the receiver-side transport-layer and application-layer observations to make a proper ABR decision through the use of deep reinforcement learning (DRL)-based model. 
(3) We have extensively evaluated \textit{Anableps} on a real-world trace-driven testbed using a variety of video contents under a corpus of real-life network traces. The results prove the effectiveness of our proposed \textit{Anableps}.

\section{Related Works}

\textbf{Bitrate prediction for VBR-encoded videos:} In the past, several works~\cite{pre_2, pre_3} were developed to predict VBR video traffic to improve the effectiveness of network management (e.g., dynamic bandwidth allocation). However, none of them uses predicted traffic to facilitate bitrate adaptation in  RTVC systems. Besides, a single fixed value is predicted to represent the VBR video traffic, which makes it easily fall into a high prediction error given the diversity of video content.

\textbf{State-of-the-art ABR algorithms:} Existing ABR algorithms can be mainly categorized as rule-based or learning-based algorithms. Rule-based algorithms usually adapt bitrates based on a fixed rule of certain network states, such as loss~\cite{abr_loss}, delay~\cite{abr_delay}, or both of them~\cite{gcc}. For example, GCC~\cite{gcc}, a widely-used congestion control algorithm, estimates the bandwidth based on the statistics of delay interval and packet loss rate, and immediately adapts the encoding/sending bitrate according to the estimation. Recently, learning-based approaches attract widespread attention for their superior performances. For example, ARS~\cite{ars} uses DRL to train a neural network for bitrate decisions according to transport-layer and application-layer states. Instead of maximizing the video bitrate, QARC~\cite{qarc} aims at optimizing the perceptual video quality, for which it introduces the prediction of video multi-method assessment fusion (VMAF) scores into input states. Furthermore, Concerto~\cite{conerto} optimizes video bitrate control based on the network information both from the transport-layer and codec-layer using imitation learning, OnRL~\cite{onrl} closes the simulation-to-reality gap by online reinforcement learning based on the transport-layer network information only, Loki~\cite{loki} improves the robustness of learning-based model by integrating it with a rule-based algorithm at the feature level, but still based on the transport-layer network information.
To sum up, all these existing algorithms overlook the video bitrate fluctuation issue which is generally existed but apparently more noticeable for VBR-encoded video, yielding inferior performances in RTVC sessions using VBR-encoding mode. 

\begin{figure*}[htbp]
\setlength{\abovecaptionskip}{0.cm}
\subfigure[system architecture]{
\centering
\includegraphics[width=0.41\textwidth]{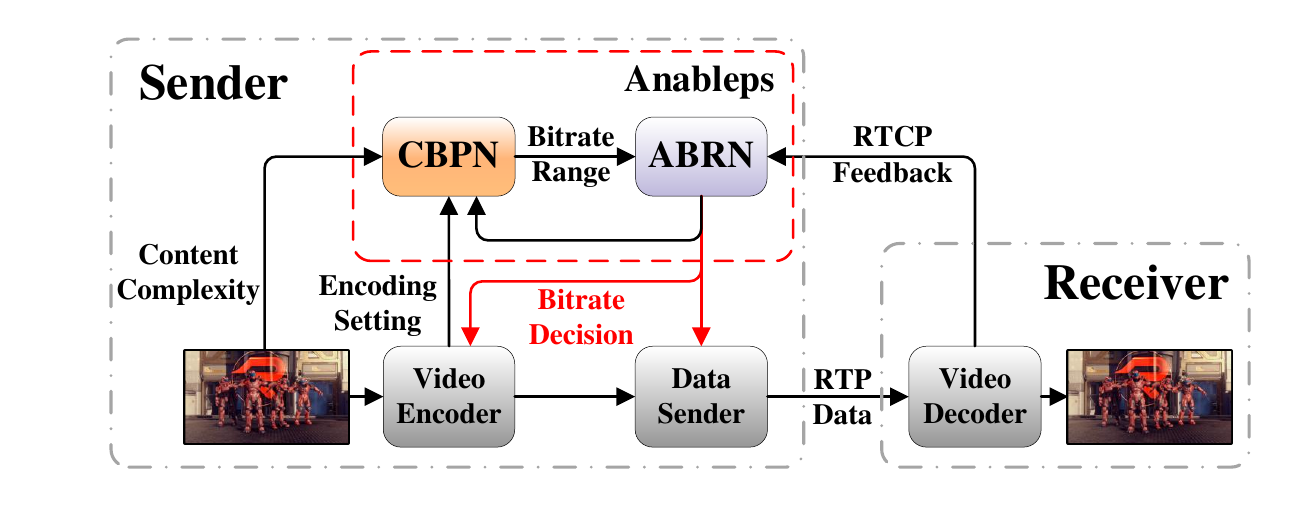}
\label{sfig:sys}
}
\subfigure[CBPN architecture]{
\centering
\includegraphics[width=0.27\textwidth]{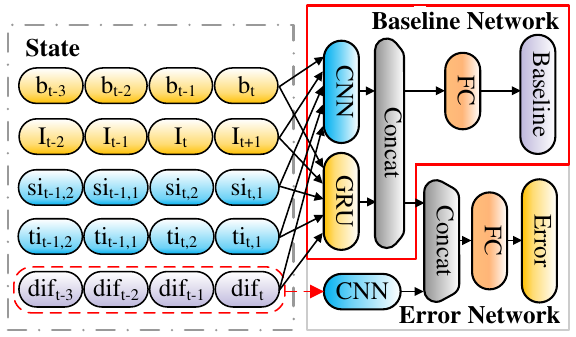}
\label{sfig:CBPN}
}
\subfigure[ABRN's actor network architecture]{
\centering
\includegraphics[width=0.26\textwidth]{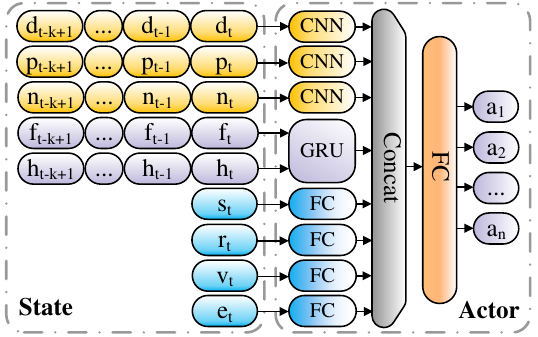}
\label{sfig:ABRN}
}
\caption{\textbf{\textit{Anableps}} leverages two cascaded neural models, i.e., CBPN and ABRN, to make proper bitrate decisions for the video encoder to combat the network dynamics and VBR-encoding-induced bitrate fluctuations. The CBPN uses the content complexity, encoding settings, and bitrate decisions in the past to predict near-future bitrate range; and the ABRN then uses this bitrate range and receiver-side observations to decide the target bitrate of upcoming time slot for encoding.} 
\label{fig:anableps}
\vspace{-0.5cm}
\end{figure*}

\section{Method}
\subsection{System Architecture}\label{AA}
The proposed \textit{Anableps} is shown in Fig.~\ref{sfig:sys}. 
On the sender side, instantaneously captured scenes are encoded using VBR mode in real-time to generate RTP-compliant packets for network delivery while at the receiver, the compressed video stream is obtained and fed into the video decoder for playback. At the same time, receiver-side observations including transport-layer network condition and application-layer playback status are sent back to the sender through RTCP packets. 
As seen, \textit{Anableps} makes proper bitrate decisions for adaptation to both the underlying network dynamics and video-related bitrate fluctuations. 

This work places the \textit{Anableps} at the sender side for examination. It consists of two cascaded modules: CBPN and ABRN, where the CBPN predicts the bitrate range of future video packets based on the content complexity, encoding settings, and bitrate decisions in the past and the ABRN leverages the DRL to make bitrate decisions to guide the video encoding at the sender for an upcoming time slot using receiver-side observations and bitrate range prediction from the CBPN. 

\vspace{-0.1cm}
\subsection{Compressed Bitrate Prediction Network (CBPN)}
Sender-side information, a.k.a., content complexity, encoding settings, and bitrate decisions, from past video frames (see Fig.~\ref{sfig:sys}) are input to a neural network-based CBPN, shown in Fig.~\ref{sfig:CBPN}, for the prediction of video compression bitrate range of video frames in the upcoming time slot.

\textbf{State:} Video compression bitrate is closely related to the spatiotemporal complexity of underlying content~\cite{vbr_complexity}. 
Here we use SI (Spatial Index) and TI (Temporal Index)~\cite{itu-ti-si} to represent the spatial and temporal complexity of the video content respectively. Four SIs ($\overrightarrow{si}_t$) and four TIs ($\overrightarrow{ti}_t$) sampled uniformly in the past two seconds are taken as the input states. To reduce the computational complexity, we downsample the raw video in both spatial (resolution from 1920$\times$1080 to 192$\times$108) and temporal (frame rate from 25FPS to 4FPS) dimensions to compute the SI and TI.

On the other hand, the placement of the I frame during video encoding also greatly affects the bitrate. For instance, we often observe the shape increase of bitrate when inserting an I frame. Nevertheless, in practice, the use of the I frame can support random access during the streaming and prevent error propagation introduced by the network deterioration. We then introduce the indicator ($\vec{I}_t$) which signals whether I frame is used in the past 3 seconds and for the next second. The future existence of the I frame is known when we preset the group of picture (GoP) setting for video encoding. 

Further, it is difficult or even practically impossible to have a perfect match between the target bitrate ($\vec{b}_t$) from the ABR model, and the actual bitrate generated by the video encoder. To make the CBPN thoroughly understand the video bitrate variations,  we include both the $\vec{b}_t$ and the differences ($\overrightarrow{dif}_t$) between the target bitrate and the corresponding actual bitrate in the past 4 seconds, into the input for learning. In summary, the state is defined as $S_t=(\vec{b}_t, \vec{I}_t, \overrightarrow{si}_t, \overrightarrow{ti}_t, \overrightarrow{dif}_t)$.

\textbf{Output:} As for latency-sensitive RTVC applications, there is no sufficient time for the rate control in the video encoder to catch up with the target bitrate perfectly. As a result, the actual video encoding bitrate may vary greatly on account of numerous factors like SI/TI-related scene dynamics and codec setting (e.g., I frame ratio). Thus the output of CBPN at a given time is a vector that consists of a baseline value ($v$) and an error offset ($e$), by which we finally obtain the prediction of bitrate range $[v-e, v+e]$ for future frames.

\textbf{Neural network architecture:} The proposed CBPN is implemented using a bounded network \cite{pre_delay}. The state matrix is first input into a convolutional neural network (CNN) layer with 32 filters, each of size 5 with stride 1, and a gate recurrent unit (GRU) layer with 32 neurons concurrently for feature extraction. These features are then merged through a concatenation layer and passed into a fully connected (FC) layer to predict the baseline value of future bitrate (baseline network in Fig.~\ref{sfig:CBPN}). 
To best mimic the practical scenario, we further introduce an extra CNN layer to extract features from $\overrightarrow{dif}_t$ which is then combined with those merged features to estimate the error of predicted bitrate (error network in Fig.~\ref{sfig:CBPN}). During the training phase, the target bitrate is randomly generated. We first train the baseline network and then freeze it to train the error network.
\subsection{Adaptive Bitrate Network}

The bitrate range from the CBPN is then used as one input state of ABRN for bitrate decision
as in Fig.~\ref{sfig:ABRN}. The ABRN relies on a neural network to  map input states to the next bitrate decision. And we train this neural network using A3C~\cite{mnih2016asynchronous}, a state-of-the-art actor-critic DRL method. The designs of ABRN's state, neural network architecture, action, and reward, are briefed as follows.

\textbf{State:} The state at step $t$ is defined as $S_t$ = ($v_t$, $e_t$, $s_t$, $r_t$, $\vec{d}_t$, $\vec{p}_t$, $\vec{n}_t$, $\vec{f}_t$, $\vec{h}_t$). $v_t$ and $e_t$ are respectively the baseline value and the error value of the next compressed bitrate predicted by CBPN, which are used to inform the video bitrate fluctuation for ABRN. Current sending bitrate ($s_t$) and receiving bitrate ($r_t$), together with the RTT ($\Vec{d_t}$), the packet loss rate ($\Vec{p_t}$), and the negative acknowledgment (NACK) received ($\Vec{n_t}$), are combined to reflect the transport-layer network condition. In addition, the lost frame rate ($\Vec{h_t}$) and average frame delay ($\Vec{f_t}$) are introduced to represent the application-layer playback status. In these state inputs, we consider the vector forms of RTT, packet loss rate, NACK, lost frame rate, and average frame delay in the past $k=6$ seconds to learn from experience.

\textbf{Neural network architecture:} ABRN uses two neural networks, i.e., actor network and critic network, to train its ABR policy. Note that only actor network is required to make bitrate decisions during the inference. In the actor network, the state inputs including the baseline and error value of the predicted bitrate and the sending and receiving bitrate are passed into their respective FC layers for feature extraction. Additionally, 3 parallel 1D-CNN layers with 128 filters are placed to respectively extract the features of the round-trip time (RTT), packet loss rate, and NACK received, and another GRU layer with 128 neurons is used to extract the features of lost frame rate and average frame delay. Results from these layers are then aggregated in a 128-neuron hidden layer with the softmax function to output the \textit{policy} (i.e., action probability), as shown in Fig.~\ref{sfig:ABRN}. The critic network uses the same network structure, except for a linear neuron for the final output \textit{value} without activation.

\textbf{Action:} The output of actor network is a vector that represents the probability distribution of a set of actions, and their sum is 1. Different from the absolute bitrate actions in existing ABR approaches, the action set used in ABRN is formulated as \{X, -400, 0, 200, 400, 600\}kbps, which represents the change relative to the previous bitrate decision. This design not only benefits the smoothness of bitrate/quality changes but also enlarges the range of bitrate adaptation without increasing the size of the action space. A special action ``X'' is set to handle the situation of a sudden bandwidth drop, and the next bitrate can be calculated by $b_{t+1} = b_{t}\times(1-p_{t})$. $b_{t+1}$ denotes the new bitrate target for encoding and sending, and $b_{t}$ is the previous bitrate decision at time $t$. 

\textbf{Reward:} The goal of RL is to maximize the expected cumulative discounted reward. In order to learn better ABR algorithms through reinforcement, the reward function is defined as follows to signal the QoE:

\begin{equation}
\setlength{\abovedisplayskip}{4pt}
\setlength{\belowdisplayskip}{4pt}
R_{t} = \alpha m_{t} - \lambda |m_{t} - m_{t-1}| - \gamma h_{t} - \delta f_{t}.
\end{equation}
Here $m_{t}$ is the VMAF score at step $t$, which directly reflects the perceived quality of compressed video by a user. The second term penalizes changes in video quality to favor smoothness. And $h_{t}$ and $f_{t}$ are the lost frame rate and the average frame delay mentioned before. $\alpha$, $\lambda$, $\gamma$ and $\delta$ are configurable parameters according to different user preferences. In this paper, we set $\alpha=8$, $\lambda=0.5$, $\gamma=4$ and $\delta=2$ empirically to train a general model.
\vspace{-0.1cm}
\section{Evaluation}
\subsection{Methodology}
\textbf{Testbed:} We built a testbed based on the open-source project Razor\footnote{Razor is a WebRTC-based real-time video conferencing project, and can be accessed through \url{https://github.com/yuanrongxi/razor}.} to evaluate the proposed \textit{Anableps} through practical RTVC sessions. In the testbed, two PCs (one serves as the sender, and the other as the receiver) are connected via a relay server, in which the traffic control tool is used to reproduce different network conditions following various network traces collected in the real world. During each session, a set of video sequences are encoded and transmitted sequentially. \textit{Anableps} and other state-of-the-art ABR algorithms, were deployed at the sender in order to quickly respond to the network dynamics.


\textbf{Datasets:} For the video dataset, we used a total of 57 video clips from the YouTube UGC video dataset~\cite{ugc_video} with different categories, including sports, lecture, gaming, etc. Among them, 47 video clips were randomly selected as the train set, and the remaining clips as the test set. For the network dataset, we collected network traces extensively from several popular public datasets following the principle proposed in~\cite{vonikakis2017probabilistic}, including (1) Oboe~\cite{oboe}: a set of throughput traces from real user sessions on wired, WiFi and cellular connections; 
(2) NYU-METS~\cite{NYU}: an LTE mobile bandwidth dataset measured in New York City metropolitan areas; (3) Norway~\cite{norway}: a 3G/HSDPA network trace dataset collected when riding on vehicles; (4) FCC~\cite{fcc}: a broadband network trace dataset. 
Note that We unified the time granularity of these traces at the 0.5-second interval for easier processing. Finally, we get a 30-trace dataset with uniform distribution of mean value, standard deviation, and coefficient of variation of network bandwidth, which well covers a wide range of network conditions. We randomly divide these traces into 80\% (24) training ones and 20\% (6) testing ones. 

\textbf{Video encoding settings:} We encoded these videos using the ultrafast preset, x264 encoder in low-delay P coding and VBR rate control modes. Other detailed configurations were set as follows: resolution=1920x1080, frame rate=25, quantization parameter (QP) ranging from 18 to 50, GOP=125, and target bitrate truncated into $[300,6100]$kbps. Following the suggestion in~\cite{apple}, we configured the $vbv\_max\_bitrate$ as twice the average bitrate $i\_bitrate$.

\begin{figure}
\centering
\setlength{\abovecaptionskip}{0.cm}
\includegraphics[width=0.42\textwidth]{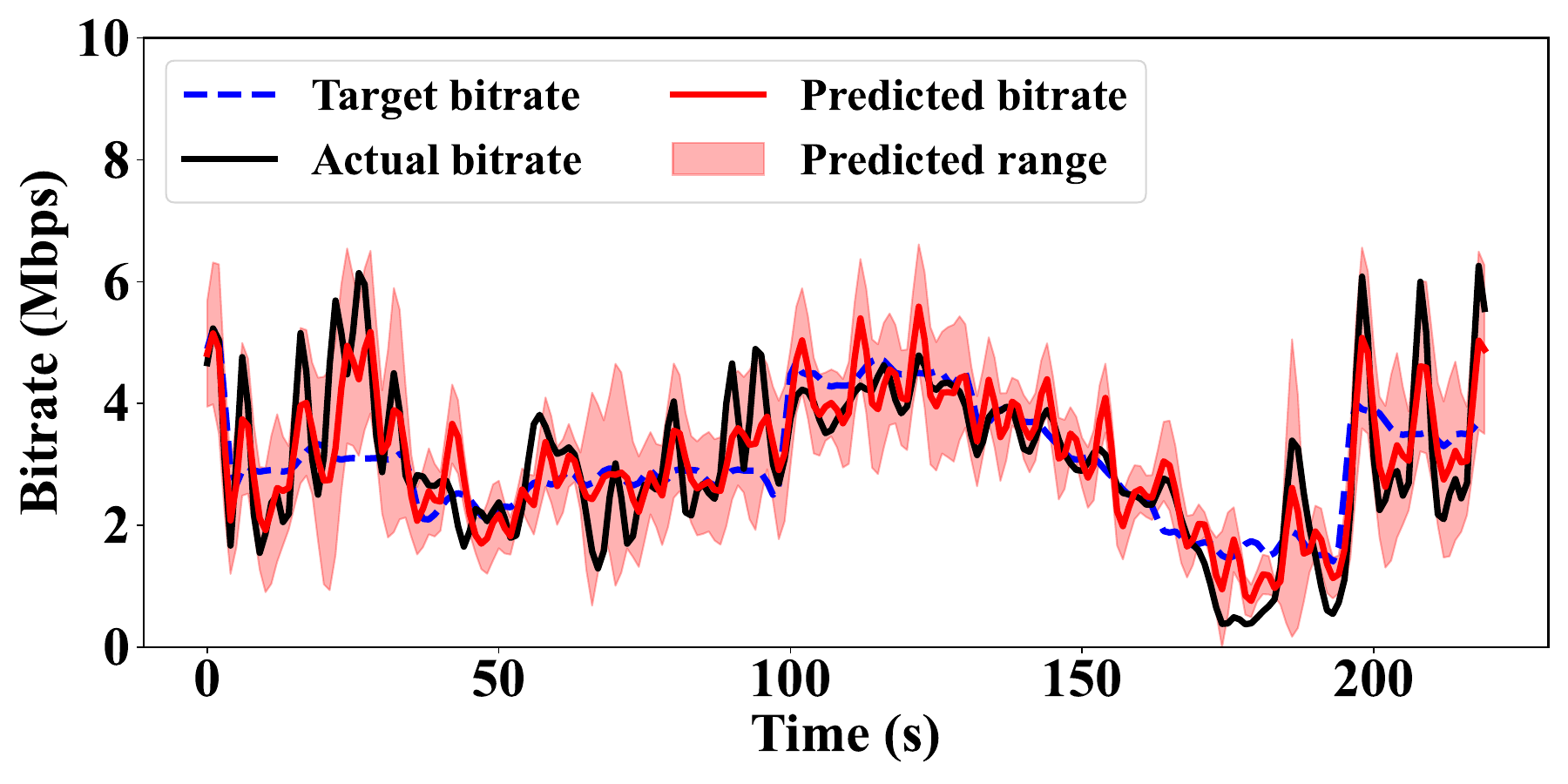}
\caption{Comparing the prediction accuracy of CBPN with the one without using bitrate range. The result shows that the predicted bitrate in a range can better cover the future bitrate fluctuation.} \label{value_error}
\vspace{-0.2cm}
\end{figure}

\begin{table}[t]
  \begin{center}
    \caption{The performance of CBPN using different input lengths.}
    \label{tab:length}
    \begin{tabular}{c|c|c|c} 
      \hline
      Length & 2 & \textbf{4} & 6 \\
      \hline
      MAD  $\downarrow$ & 0.241 & \textbf{0.218} & 0.247 \\
      PCC $\uparrow$ & 0.802 & \textbf{0.838} & 0.821 \\
      CR $\uparrow$ & 80.33\% & \textbf{85.68\%} & 82.5\% \\
      \hline
    \end{tabular}
  \end{center}
  \vspace{-0.55cm}
\end{table}

\begin{figure*}[htbp]
\centering
\setlength{\abovecaptionskip}{0.cm}
\subfigure[VMAF and sending bitrate]{
\includegraphics[width=0.38\linewidth]{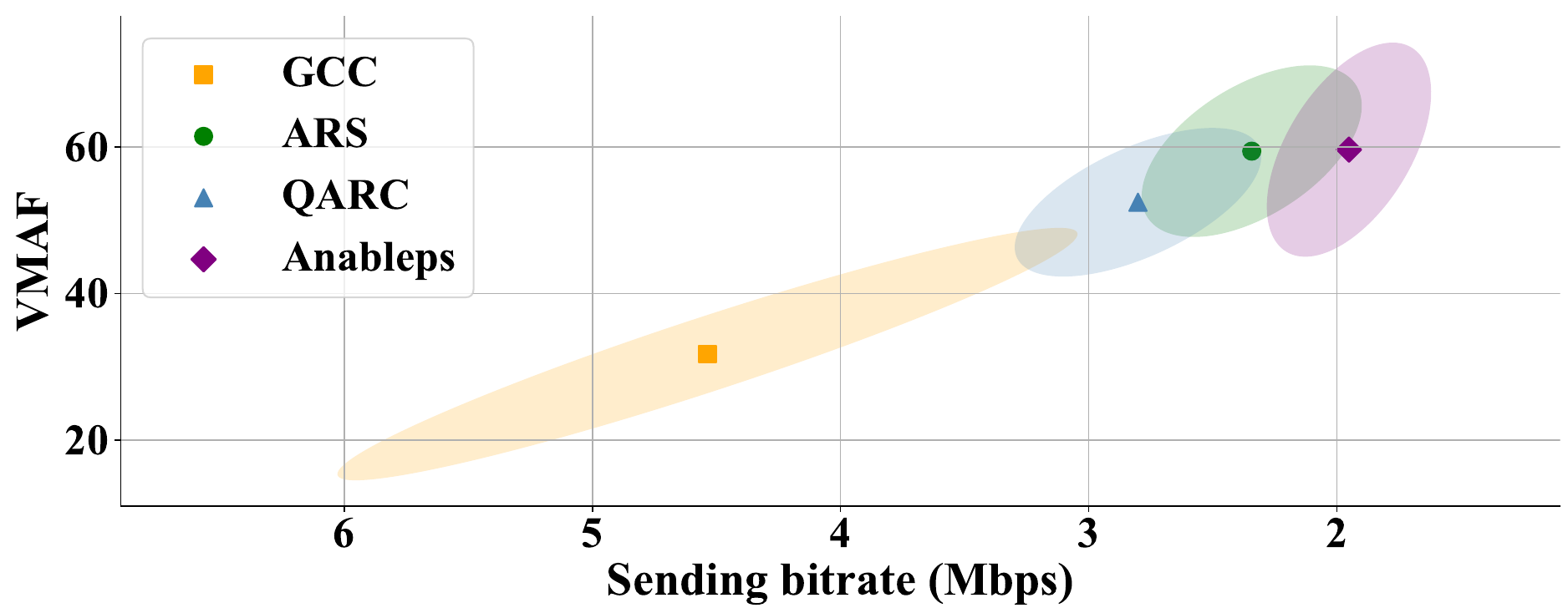}
\label{sfig:gain}
}
\hspace{1.5cm}
\centering
\subfigure[stalling ratio and frame delay]{
\includegraphics[width=0.45\linewidth]{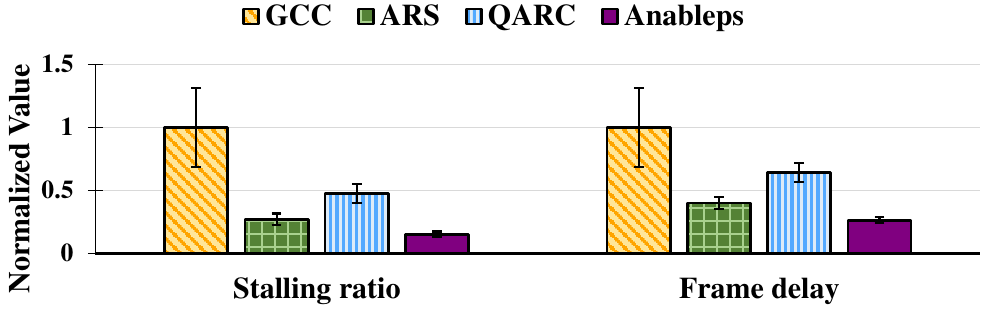}
\label{sfig:gain2}
}
\caption{Comparing \textit{Anableps} with other ABR algorithms on the metrics of VMAF, sending bitrate, stalling ratio, and frame delay. The ellipses/error bars span $\pm$ one standard deviation from the average.}
\label{fig:gains}
\vspace{-0.6cm}
\end{figure*}

\begin{figure}[htbp]
\centering
\setlength{\abovecaptionskip}{0.cm}
\includegraphics[width=0.9\linewidth]{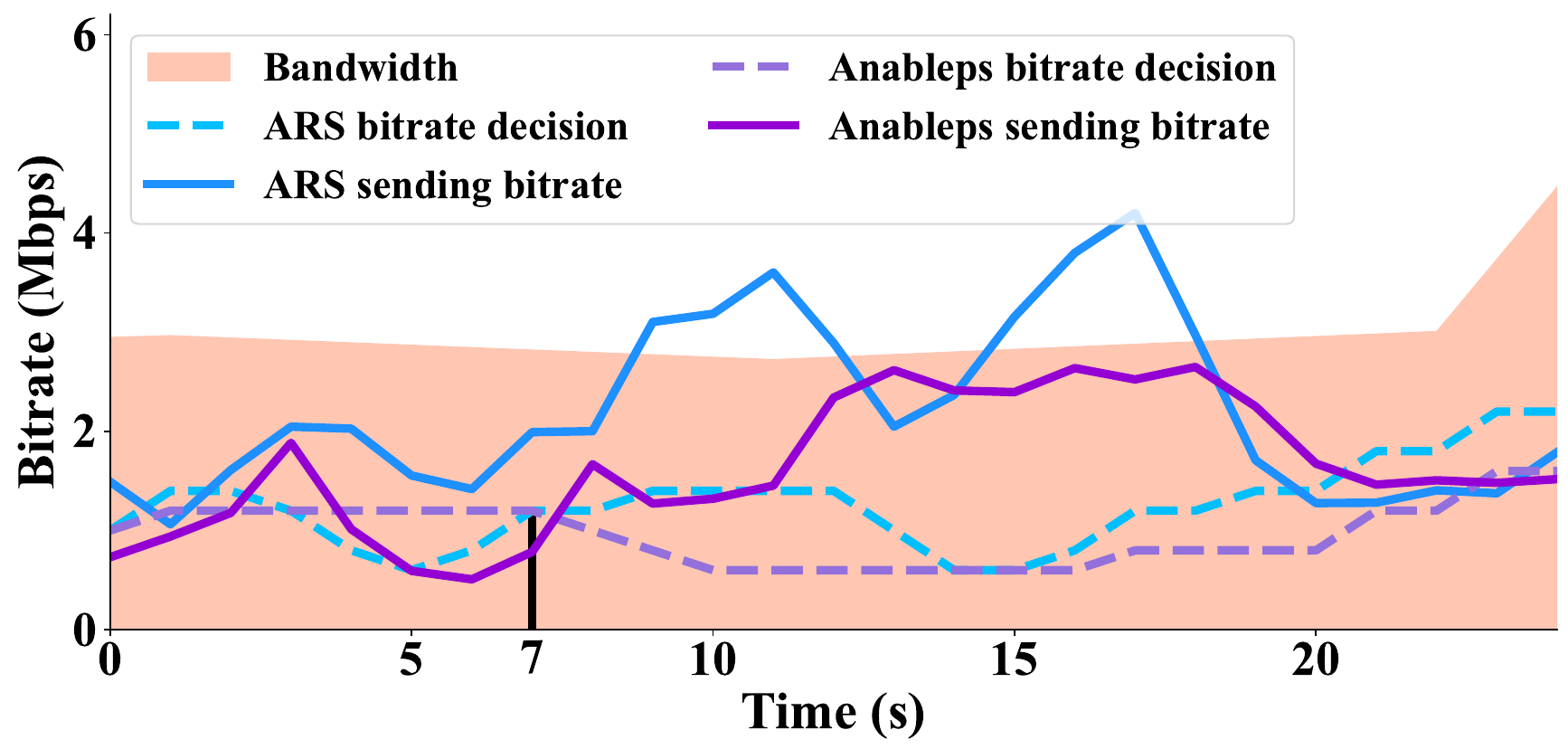}
\caption{Comparing the bitrate decision process over time using \textit{Anableps} with that using ARS in an RTVC session under a randomly selected network trace.} 
\label{fig:ars_ana}
\vspace{-0.45cm}
\end{figure}

\textbf{Baseline algorithms:} We compare \textit{Anableps} with the following state-of-the-art ABR algorithms for RTVC applications:
(1) {Google congestion control (GCC),} a heuristic method that adapts the encoding and transmission bitrate based on the currently available bandwidth, which is estimated mainly using the packet loss rate and the filtered delay intervals.
(2) {Adaptive Real-time Streaming (ARS),} a DRL-based algorithm that considers playback status into the state to make bitrate decisions, aiming at maximizing the QoE in the cloud gaming, a typical real-time communication application.
(3) {Video Quality Aware Rate Control (QARC),} a learning-based approach that aims to obtain a higher perceptual video quality with possible lower sending rate and transmission latency. It selects bitrate for future video frames based on predicted video quality and observed network status.
\vspace{-0.1cm}
\subsection{The CBPN Performance}
To evaluate the performance of CBPN, we randomly generate another series of target bitrates (i.e., $b_t$) as the testing set, and compare the predicted bitrate output by CBPN and the actual compressed bitrate output by the video encoder. A part of the results is shown in Fig.\ref{value_error}. We find that due to the large target-actual bitrate gap induced by VBR encoding, the predicted bitrate without range fails to well describe the future bitrate fluctuation, while the predicted bitrate range from CBPN can effectively cover the future actual bitrate. To quantitatively evaluate the CBPN performance, we introduce the cover ratio (CR) metric proposed in \cite{pre_delay} to measure the prediction accuracy of CBPN, which is defined as follows:
\begin{equation}
\setlength{\abovedisplayskip}{4pt}
\setlength{\belowdisplayskip}{4pt}
\frac{1}{n}\sum_{i=1}^{n}C_{i}, where~C_{i}=\begin{cases}1,~\hat{y} \in [v-e,v+e]
 \\0,~\hat{y} \notin [v-e,v+e]
\end{cases}
\end{equation}
where ${\hat{y}}$ represents the actual compressed bitrate, and ${n}$ is the size of test dataset. The experimental results on the testing set show that CBPN finally gets a CR of 85.68\%, which is accurate enough to predict the actual bitrate of upcoming video frames. We believe that it is conducive to better ABR decisions, which will be verified in the next subsection.

Additionally, the timing window length of the input state in the CBPN is determined based on performance evaluation. The result is provided in Table~\ref{tab:length}. Besides CR, two more metrics including mean absolute difference (MAD) and Pearson correlation coefficient (PCC) are introduced to compare the prediction performances using the different lengths of the timing window. 

\vspace{-0.2cm}
\subsection{\textit{Anableps} vs. Other ABR Algorithms}
We run the four considered ABR algorithms in the testbed. During each session, we record the results on several QoE metrics, including the sending bitrate, stalling ratio, average frame delay, and video quality, to evaluate their performances. The sending bitrate metric represents the bandwidth consumption of an RTVC session. The stalling ratio indicates the ratio of stalling
events that playing video frame rate drops below 12 fps~\cite{onrl}. And the average frame delay measures the interaction delay between the two ends. In this work, we use the average VMAF score of video frames that are received and finally played at the receiver to measure the video quality.

As shown in Fig.~\ref{fig:gains}, the proposed \textit{Anableps} achieves significant improvements almost on every QoE metric. Compared with the GCC, ARS, and QARC algorithms, \textit{Anableps} not only improves the VMAF on average by 87.92\%, 0.35\%, and 13.68\% respectively, but also reduces the bandwidth consumption by 56.99\%, 16.73\%, and 30.36\%. Meanwhile, \textit{Anableps} greatly reduces the stalling ratio by 84.85\%, 44.30\%, and 68.10\%, and lowers the interaction delay by 73.61\%, 33.97\%, and 58.97\%. This is because \textit{Anableps} bridges the target-actual video bitrate gap induced by VBR encoding when performing bitrate adaptation, while other ABR algorithms fail to consider this issue. On the other hand, \textit{Anableps} gains the minimum standard deviations of sending bitrate, stalling ratio, and frame delay at acceptable quality variations for different videos. This result reveals that \textit{Anableps} can achieve consistent good ABR performances for different videos by learning their content features. 

Among these existing algorithms, GCC is the worst-performing one, with the highest sending bitrate, highest stalling ratio, and longest frame delay. We conjecture that GCC is essentially a congestion control algorithm, which aims to maximize the bandwidth utilization instead of optimizing the QoE. QARC performs better than GCC, as it is designed to obtain a higher perceptual video quality with possibly lower sending rate and transmission latency. But it fails to consider the playback fluency, a significant factor for the QoE, making its performance inferior. Except for the proposed \textit{Anableps}, the best performance is achieved by using the ARS algorithm, because it takes the stalling into additional consideration when making bitrate decisions. Nevertheless, ARS still suffers from frequent overshooting events and low QoE when encountering violent bitrate fluctuations, which is a common phenomenon in RTVC sessions using VBR-encoded video.

To evaluate the instant ABR performance of \textit{Anableps}, we also plot sequential bitrate decisions over time in an RTVC session under a randomly selected network trace, as shown in Fig.~\ref{fig:ars_ana}. The result of the best-performing ARS in the baseline algorithms is also plotted for comparison. As we can see, \textit{Anableps} perceives the potential threat of overshooting induced by the compressed bitrate fluctuation at the 7th second, and decides on a lower target bitrate for encoding, making the final sending bitrate smaller than the available bandwidth. By contrast, ARS fails to consider and predict such video bitrate fluctuation, so it continues to increase the target bitrate and inevitably harvests a prolonged overshooting event, which severely degrades the QoE in terms of frequent stalling and high interaction delay. This result further verifies \textit{Anableps}'s capability to adapt bitrate to both underlying network dynamics and video bitrate fluctuations.

\subsection{\textit{Anableps} Deep Dive}
In order to understand the contribution of each module of \textit{Anableps} (i.e., CBPN and VBRN) to ABR performance improvement, we conduct an experiment to compare the performances of default \textit{Anableps}, \textit{Anableps} without using CBPN (denoted as \textit{Anableps-s}) and \textit{Anableps} without using the error value of predicted bitrate (denoted as \textit{Anableps-c}). In \textit{Anableps-s}, CBPN is disabled, and the state components related to bitrate prediction (i.e., the baseline and error values of predicted bitrate) in ABRN are also disabled. While in \textit{Anableps-c}, CBPN is enabled, and the error value of the predicted bitrate is disabled for ABRN. We retrain \textit{Anableps-s} and \textit{Anableps-c} using the same training algorithm as the default \textit{Anableps} in the same environment. Then the well-trained \textit{Anableps-s} and \textit{Anableps-c} are also evaluated in the aforementioned testbed.

Table~\ref{tab:deep_dive} lists the performance improvements of default \textit{Anableps}, \textit{Anableps-s}, and \textit{Anableps-c} on all considered QoE metrics over the best-performing baseline ARS. We find that \textit{Anableps-s} with the ABRN alone could greatly improve the ABR performance, i.e., reducing the bandwidth consumption by 10.62\%, the stalling ratio by 33.99\%, and the interaction delay by 27.53\%, while at the same time improving the video quality by 0.57\%. This verifies the effectiveness of ABRN. \textit{Anableps-c} further improves the performance by introducing the baseline value of predicted bitrate to facilitate ABR decision. Taking the ARS algorithm as the anchor, default \textit{Anableps} can obtain additional performance gains, with the reduction of bandwidth consumption by 6.11\% and 4.53\%, stalling ratio by 10.31\% and 9.85\%, and frame delay by 6.44\% and 5.94\% at the similar video quality, in comparison to \textit{Anableps-s} and \textit{Anableps-c} respectively. These results indicate the necessity of considering the bitrate fluctuation induced by VBR encoding when designing ABR approaches and the effectiveness of CBPN.

\begin{table}[t]
\vspace{-0.3cm}
  \begin{center}
    \caption{The performance improvements of default \textit{Anableps}, \textit{Anableps-s}, and \textit{Anableps-c} over the best-performing baseline ARS.}
    \label{tab:deep_dive}
    \begin{tabular}{c|c|c|c} 
      \hline
      Metric & \textit{Anableps-s} & \textit{Anableps-c} & \textit{Anableps} \\
      \hline
      VMAF $\uparrow$ & \textbf{+0.57\%} & +0.10\% & +0.35\% \\
      Sending bitrate $\downarrow$ & -10.62\% & -12.20\% & \textbf{-16.73\%} \\
      Stalling ratio $\downarrow$ & -33.99\% & -34.45\% & \textbf{-44.30\%} \\
      Frame delay $\downarrow$ & -27.53\% & -28.03\% & \textbf{-33.97\%} \\
      \hline
    \end{tabular}
  \end{center}
  \vspace{-0.6cm}
\end{table}

\section{CONCLUSION}
In this paper, we present and evaluate \textit{Anableps}, a learning-based bitrate adaptation approach for VBR-encoded video transmission in the real-time communication system. Unlike existing ABR algorithms, \textit{Anableps} considers and bridges the target-actual video bitrate gap induced by VBR encoding, by predicting the range of future bitrate and further making subsequent bitrate decisions based on sender-side bitrate prediction and receiver-side observations collected from the past. Through extensive trace-driven field tests on a testbed, the results report that \textit{Anableps} outperforms existing ABR algorithms for the RTVC scenario.


\bibliographystyle{IEEEtran}
\bibliography{bibtex}

\vspace{12pt}

\end{document}